\documentclass[authoryear,review]{elsarticle}
\usepackage{mathrsfs,ifthen}

\usepackage{amssymb}

\usepackage[usenames]{color}

\usepackage{graphicx,subfigure}
\graphicspath{{./}{./Figs/}}

\usepackage[CJKbookmarks,bookmarks=true]{hyperref}
\hypersetup{colorlinks, linkcolor=blue, citecolor=blue, urlcolor=black,%
    plainpages=true, bookmarksopen=false,%
    pdfhighlight=/P, 
    pdfauthor={Shan Mei<MeiShan.ann@gmail.com>},%
    pdftitle={Simulating City-level Airborne Infectious Diseases},%
    pdfsubject={LaTeX, TeX},%
    pdfkeywords={LaTeX, TeX, Typesetting},%
    pdfstartview=FitH, 
    pdfview=FitH,
    pdfpagemode=UseOutlines, 
}

\usepackage{algorithm, algorithmic}

\usepackage{tabularx,multirow,hhline}

\journal{Computers, Environment and Urban Systems}

\begin{document}

\begin{Large}\textbf{Simulating City-level Airborne Infectious Diseases}\end{Large}

\
\

Shan Mei$^{1,2,*}$, Xuan Zhou$^{1}$, Yifan Zhu$^{1}$, Zhenghu Zu$^{3}$, Tao Zheng$^{3}$, A.V. Boukhanovsky$^{4}$, P.M.A Sloot$^{2,4,5}$

\

\

$^{1}${National University of Defense Technology, P. R. China}

$^{2}${Computational Science, University of Amsterdam, Amsterdam, Netherlands}

$^{3}${Beijing Institute of Biotechnology, Academy of Military Medical Science, P. R. China}

$^{4}${National Research University ITMO, Russia}

$^{5}${Nanyang Technological University, Singapore}

$^{*}${Corresponding author. Telephone: (+)86-731-84573558, Fax: (+)86-731-84573535. }

\

\

Email addresses:

SM: \href{mailto:Meishan.ann@gmail.com}{Meishan.ann@gmail.com}

XZ: \href{mailto:qiangjunxingguo@163.com}{qiangjunxingguo@163.com}

YZ: \href{mailto:stephen.zhuyifan@gamil.com}{stephen.zhuyifan@gamil.com}

ZZ: \href{mailto:zzhbiot08@126.com}{zzhbiot08@126.com}

TZ: \href{mailto:zhengtao@126.com}{zhengtao@126.com}

AVB: \href{mailto:avb\_mail@mail.ru}{avb\_mail@mail.ru}

PMAS: \href{mailto:P.M.A.Sloot@uva.nl}{P.M.A.Sloot@uva.nl}

\begin{frontmatter}

\begin{abstract}
With the exponential growth in the world population and the constant increase in human mobility, the danger of outbreaks of epidemics is raising. Especially in high density urban areas such as public transport and transfer points, where people come in close proximity of each other, we observe a dramatic increase in the transmission of airborne viruses and related pathogens. It is essential to have a good understanding of the `transmission highways' in such areas, in order to prevent or to predict the spreading of infectious diseases. The approach we take is to combine as much information as is possible, from all relevant sources and integrate this in a simulation environment that allows for scenario testing and decision support. In this paper we lay out a novel approach to study Urban Airborne Disease spreading by combining traffic information, with geo-spatial data, infection dynamics and spreading characteristics.

\end{abstract}


\end{frontmatter}

\section{Introduction}
\label{Introduction}

%
%

City-level airborne epidemics is a threat to healthy living. With the exponential growth in the world population and the constant increase in human mobility, the danger of outbreaks of epidemics is raising. For example, the novel Influenza A (H1N1), also known as Human Swine Influenza/Swine Flu, spreading internationally from Mexico in 2009, has caused a serious epidemic in China. China is highly susceptible to pandemic influenza A
(H1N1) due to its big population and high residential density. According to the Ministry of Health of China, until 30th Sep 2009, the provinces in China mainland had reported 19589 confirmed cases, 14348 cured cases, 10 sever cases and a few death cases \citep{ChinaReport2009}.

In high density urban areas such as public transport and transfer points, where people come in close proximity of each other, we observe a dramatic increase in the transmission of airborne viruses and related pathogens. In order to elaborately model and simulate the airborne epidemics, the city under study needs to be modeled in detail from the infrastructural aspect. We utilize the Geographic Information System (GIS) technology to model the infrastructure of a city which might be threatened by certain epidemic attacks. GIS is a combination of database management capabilities for collecting and storing large amounts of geospatial data, together with spatial analysis capabilities to investigate geospatial relationships among the entities represented by that data, plus map display capabilities to portray the geospatial relationships in two- and three-dimensional map form \citep{Nyerges2009}. GIS facilitates storing, querying and visualizing city infrastructure including roads, regions with diverse functionality, public transportation and so forth. We also address path routing based on city transportation to capture transmissions that occur to localities, especially public transport. This is because in many developing countries such as China, the overly crowded public transportation usually escalates airborne epidemics.


On the basis of the geo-spatial information, we model a local population that dwell in a city under study and their spatio-dynamical behavior. There is growing recognition that the solutions to the most vexing public health problems are likely to be those that embrace the behavioral and social sciences as key players \citep{Mabry2008}. Human behavior plays an important role in the spread of infectious diseases, and understanding the influence of behavior on the spread of diseases can be key to improving control efforts \citep{Funk2010}. It is essential to have a good understanding of the `transmission highways' in urban areas, in order to prevent or to predict the spreading of infectious diseases. Therefore, investigating into the patterns that are relevant for social contacts, and consequent airborne virus transmissions, is of great importance.


In this study lay out a novel approach to study Urban Airborne Disease spreading by combining traffic information, with geo-spatial data, infection dynamics and spreading characteristics. We combine as much information as is possible, from all relevant sources and integrate this in a simulation environment that allows for scenario testing and decision support.


\section{Model}

\subsection{City Modeling}

We discuss city modeling from the aspects of city partitions and traffic (road and public transportation) networks.

\subsubsection{Regions and Sublocations}

In order to construct a synthetic city, we break down major metropolitan areas into \textbf{regions} and \textbf{sublocations (SLs)} that reside inside each region. Regions, or land uses in some studies, are pieces of city land serving different purposes such as agriculture, commerce, medication and education etc. Sublocations, affiliated to a specific region, represent realistic-room like space where people conduct their daily activities and have social contacts.

Each region is categorized into types of agriculture, residence, hospital, school, university and recreation etc., due to the main facilities that it provides people with, as discussed in references \citep{Valle2006,Del2007,Yang2007}. In this study we exclusively consider 6 types of regions -- \textbf{housing (HR)}, \textbf{office (OR)}, \textbf{school (SR)}, \textbf{university (UR)}, \textbf{medical (MR)} and \textbf{recreational (RR)}. Such region partition asks for GIS files that comprise clearly partitioned land pieces, and these pieces can be mapped to the aforementioned 6 types of region, ignoring those regions (e.g., agricultural regions) that contribute less to the spread of diseases.


A region contains a set of sublocations of different classes. For example, a university region (UR) contains office sublocations (offices), residential sublocations (student dormitories and faculty members' home), classroom sublocations (classrooms, labs and library space), recreation sublocations (cafeteria, clubs, shops, refectories and restaurants etc.) and possibly hospital sublocations. Specifically, the recreational class includes shops, restaurants, cinemas, supermarkets and all other relevant places which provide services of recreation, relaxation or sales of life necessities. In this sutdy we classify sublocations as \textbf{housing (HS)}, \textbf{office (OS)}, \textbf{classroom (CS)}, \textbf{patient room (PS)} and \textbf{recreational (RS)}. Table \ref{regions and sublocations} lists the classifications of regions and secondary sublocations in detail.

\begin{center}
\textbf{TABLE 1}
\end{center}

Additionally, each sublocation is characterized as being either indoor or outdoor, conveying different transmission probabilities of viruses inside the space. For many airborne viruses, outdoor conditions such as sunshine, heat, wind blow and air circulation can lower the infection probability between the infected and the susceptible.

\subsubsection{The Road and Public Transportation Networks}

The whole city traffic routes are modeled as a road network (RN) and a public transportation network (PTN).

\begin{figure}[htbp]
\begin{center}
\includegraphics[width=200 bp]{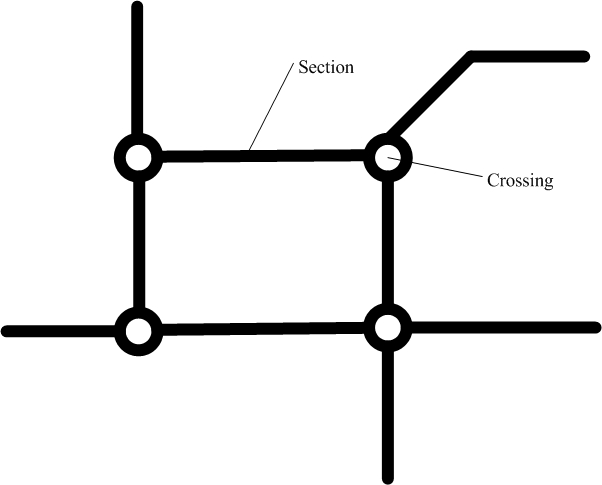}
\caption{Road Network}\label{Roads}
\end{center}
\end{figure}

The assemblage of roads in a city can be mapped to a road network (RN). Roads, as the transport infrastructure of a city, are composed of road sections and crossings. We build up the road network, denoting crossings by nodes and sections by edges, as shown in Fig. \ref{Roads}. Each edge, either a straight-line or a poly-line, stands for a head-tail combination of realistic sections. Edges can be 1-directional or 2-directional, indicating that they are corresponding to one-way or double-way road sections, respectively. A crossing joins several road sections together. The number of sections (usually 2, 3 or 4) that a crossing links to indicates the connectivity of the crossing. In this way, the whole city roadway can be mapped to a complex network \footnote{In the context of network theory, a complex network is a network with non-trivial topological features that mostly do not occur in simple networks such as lattices \citep{Newman2003}.} of vast nodes and edges. The degree of each node shows the number of neighboring sections that this crossing connects with. Obviously, the degree of nodes in this road network is greater than or equal to 1.

\begin{center}
\textbf{FIGURE 1}
\end{center}

\begin{figure}[htbp]
\begin{center}
\includegraphics[width=200 bp]{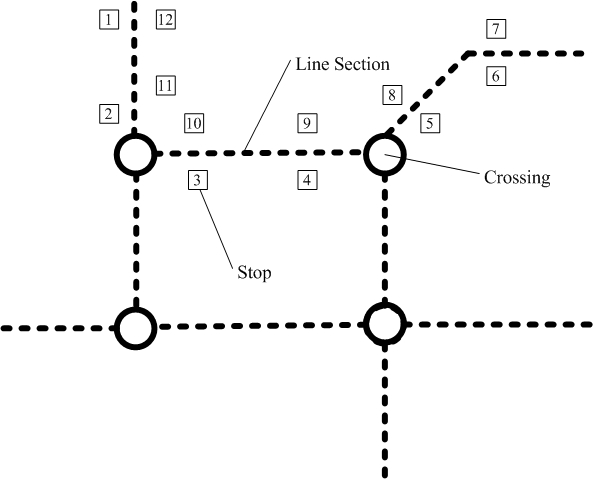}
\caption{Public Transportation Network}\label{PTN}
\end{center}
\end{figure}

The assemblage of public transportation routes can be mapped to a public transportation network (PTN) in terms of lines and stops. The public transportation consists of many bus/tram and metro/train lines along which buses/trams and metros/trains operate frequently during day time. Buses/trams and metros/trains depart every few minutes from the starting stop of a line and move towards the destination stop. People get on or off at each stop alongside each line. A number of lines join at a stop to make transfer. Every line is composed of head-tail line sections. Similar to a crossing in the roadway, a line crossing joins several line sections together. Therefore, we construct the PTN by denoting crossings by nodes and sections by edges, while stops are scattered along both sides of each edge, as shown in Fig. \ref{PTN}. The route of a bus line in one direction passes six bus stops (displayed in square) No. 1-6 alongside three line sections, while it in the other direction passes stops No. 7-12 alongside the same three line sections.


\begin{center}
\textbf{FIGURE 2}
\end{center}


\subsubsection{Travel Routing}
\label{Travel Routing}

People travel within a city by foot, car, taxi, bike and public transportation. We discuss their travel routing by utilizing the previously constructed RN and PTN. For simplicity, we focus on the mobility of people inside the city under study, leaving out of consideration the relatively less common commutes between cities.

\begin{figure}[htbp]
\begin{center}
\includegraphics[width=200 bp]{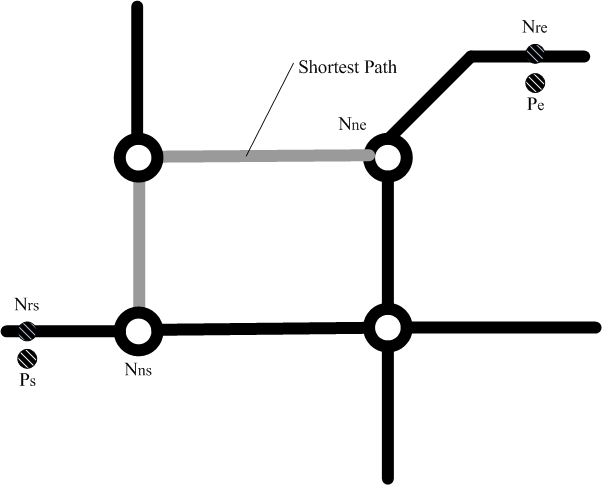}
\caption{Travel Routing for the Shortest Path}\label{SP}
\end{center}
\end{figure}

For travels by means of other than public transportation, we consider travel routing in two ways. On the one hand, people move along a straight line connecting the start-point ($P_s$) and the end-point ($P_e$) for very short distances, say, when $Distance(P_s,P_e)\leqslant$ 3 km. So $Path = L(P_s,P_e)$ where $L$ denotes the line between the two points. On the other hand, people move along the shortest or feasible paths in the RN for longer distances. The fundamental theoretic achievements in the field of complex networks can help seek the shortest or feasible paths, for example by using the so-called Dijkstra algorithm for node-to-node shortest path computation. Thus, if $Distance(P_s,P_e)>$ 3 km, the travel routing result will be the combination of 5 parts (see Fig. \ref{SP}), i.e., $Path = L(P_s,N_{rs})+L(N_{rs},N_{ns})+SP(N_{ns},N_{ne})+L(N_{ne},N_{re})+L(N_{re},P_{e})$, where $N_{rs}$ is the nearest point on road (edges in RN) to $P_s$, $N_{ns}$ is the nearest node (in RN) to $N_{rs}$, $SP$ computes the shortest path (plotted in grey) between two points in RN ($N_{ns}$ and $N_{ne}$ in this case), $N_{re}$ is the nearest point on road to $P_e$ and $N_{ne}$ is the nearest node to $N_{re}$. Additionally, the $SP$ result can be substituted by other feasible paths if traffic avoidance needs to be considered.

\begin{center}
\textbf{FIGURE 3}
\end{center}

\begin{figure}[htbp]
\begin{center}
\includegraphics[width=350 bp]{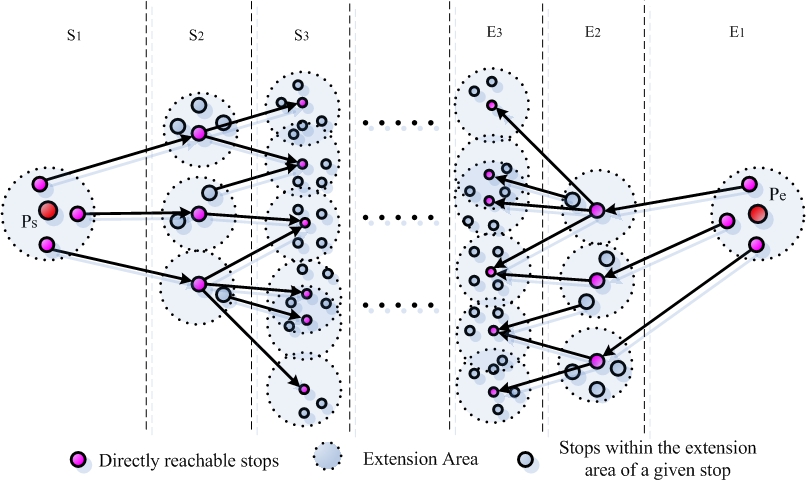}
\caption{Travel Routing by Public Transportation}\label{PT}
\end{center}
\end{figure}

Traveling by public transportation further complicates routing so that we design a breadth-first algorithm as illustrated in Fig. \ref{PT}. Looking for the shortest paths in PTN unnecessarily solve the routing problem because buses/metros move along predefined stop-by-stop lines instead of shortest paths. We start with $P_s$ and $P_e$, i.e., the start-point and the end-point. In order to compute paths based on public transportation, a set of stops close to $P_s$, denoted by $S_1$, need to obtained first. We search these stops within a given distance of $P_s$, and the resulting circle with center point $P_s$ and given radius is called the \emph{extension area} of $P_s$. We define the \emph{extension operation} $Ex(stop,radius)$ as to get all the stops inside the circle with center point $stop$ and given $radius$. Thus $S_1=Ex(P_s,radius)$. In the same way we can get $E_1$, a set of close stops to $P_e$, utilizing the corresponding extension operation $Ex(P_e,radius)$. The two sets of $S_1$ and $E_1$ represent the getting-on stops and getting-off stops that people can choose for the sake of traveling from location $P_s$ to $P_e$ by public transportation.

The search operation continues iteratively. We denote the resulting set of stops from the $i$th search by $S_i$ and the radius adopted for the $i$th search by $radius_i$. In order to conduct onward searching according to predefined line information, we define the operation of obtaining directly reachable stops of a given stop as $DR(stop)$, which produces a set of stops coming right next to the given stop in all passing-by lines. As shown in Fig. \ref{PT}, for each stop in $S_1$ (three in total except $P_s$), we can get its $DR(stop)$ set. As an illustration, in Fig. \ref{PT}, each stop in $S_1$ is assumed to have only one directly reachable stop. Thus, the union of the stops inside the three extension areas (three big circles with dashed perimeter) leads to $S_2=\displaystyle \mathop{\cup}_{stop_2\in DRS}{Ex(stop_2,radius_2)}$ where $DRS=\displaystyle \mathop{\cup}_{stop_1\in S_1}{DR(stop_1)}$. Therefore, The searching results starting from both $P_s$ and $P_e$ are given in Equ. \ref{Si} and Equ. \ref{Ei}, respectively.

\begin{equation}\label{Si}
S_i= \left\{
    \begin{array}{ll}
        Ex(P_s,radius_1) & i=1 \\
        \displaystyle \mathop{\bigcup}_{stop_1\in S_{i-1}}\displaystyle \mathop{\bigcup}_{stop_2\in {DR(stop_1)}}{Ex(stop_2,radius_i)} & i\geq 2
\end{array} \right.
\end{equation}

\begin{equation}\label{Ei}
E_i= \left\{
    \begin{array}{ll}
        Ex(P_e,radius_1) & i=1 \\
        \displaystyle \mathop{\bigcup}_{stop_1\in E_{i-1}}\displaystyle \mathop{\bigcup}_{stop_2\in {DR(stop_1)}}{Ex(stop_2,radius_i)} & i\geq 2
\end{array} \right.
\end{equation}


The searching termination condition is that $\exists i,j,S_i\cap E_j\neq \emptyset$. With minimum transits firstly and less time consumption secondly taking priorities over others, we end up with the optimal routing from $P_s$ to $P_e$, invovling all the stops on the way and necessary transits. Please note that the radius parameter for each $i$th search, $radius_i$, is tunable. For example, we can set $radius_1=1 $ km to search for stops within a radius of 1 km from $P_s$ or $P_e$, and 0.05 km for the rest implying that transits between lines can only occur when the distance between 2 nearby stops is $<0.05$ km. If the routing fails, which means that the algorithm ends up with $S_i\cap E_j = \emptyset, \forall i,j$, people need to resort to other traveling means, e.g., taxi. Arguably, in a maturely developed city, the possibility of routing failure is low and bearable for our simulations.



\begin{center}
\textbf{FIGURE 4}
\end{center}


\subsection{The Synthetic Population}

In this study, each person is represented as an agent with own attributes and behavior, based on the Multi-Agent Systems theory which has proven to be suitable for modeling epidemics \citep{Epstein1996,Reynolds2000,Heylighen2001,Koopman2006,Auchincloss2008}. In order to synthesize the population in a city under study and investigate how people transmit airborne viruses and respond to epidemics, we need to outline the attributes of people with respect to epidemiological and sociological characteristics, and model people's daily behavior, the occurrence of contacts and subsequent infections.

\subsubsection{Attributes and Classification of People}


The selection of attributes of people is determined by the virological, medical, sociological and demographical data that is available to support our models. In general, we consider age, gender, susceptibility to epidemics, immunity to certain viruses, social status, infection status (susceptible, infected, infectious, treated or cured), Housing SL and Office SL etc. The Housing SL indicates the place where a person rest especially at night, including home and dormitories etc.; in contrast, the Office SL indicates the place where a person spends time on working (for adults) or studying (for students) during day time. Please note that study activities of students are also regarded as work activities. Initially assigning values to these attributes for each person depends on the actual statistical distribution or rules deduced from available data. For example, the age distribution can be obtained from national census; the distribution of the distance between one's Housing SL and Office SL for people living in a given city can be estimated based on questionnaires. Setting one's Housing SL and Office SL complies with the following procedure: (1) each person is initially attached to a Housing SL, taking into account household formation rules regarding age, gender, and family size etc.; (2) a distance value between the Housing SL and the Office SL is drawn from a certain distribution; (3) a Office SL is randomly selected on the city map at proximately a distance of the above value away from the Housing SL.
Once set, some attributes are kept constant throughout simulations, while others can change over simulation time. For instance, the infection status of a person can be set susceptible (healthy) at the beginning, and then change in response to the occurrences of certain events such as infections.

All individuals are classified according to age structure and lifestyles, as given in Table \ref{people classification}. Subsequent modeling of people's activities (in Sec. \ref{Daily Agenda}) is applied to this 5 classifications because it is believed that daily activity patterns are related to individuals' socioeconomic characteristics such as household role, lifestyle and life cycle \citep{Kulkarni2000,Yang2007}.

\begin{center}
\textbf{TABLE 2}
\end{center}

%
%
%
%
%
%
%
%
%

\subsubsection{Daily Agenda}
\label{Daily Agenda}

People conduct diverse daily activities in sublocations. The daily activities consist of working, staying at home, relaxing at different recreational places, staying in hospital after getting infected, etc. People engage in each activity in a specific class of sublocation (the classification is given in Table \ref{regions and sublocations}). Specifically, people work in Office SLs, stay home in Housing SLs, shop/do sport/enjoy entertainment in Recreational SLs, get treatment in Patient Room SLs and study in Classroom SLs. The types of activities, the corresponding places where these activities take place and the involved people classes are listed in Table \ref{activity types}, based on research reported in \citep{Valle2006,Yang2007}. Addtionally, the reader is referred to Table one in \citep{Del2007} for the distribution of average duration by activity type. For instance, the average duration of Home activity is 12 h 24 min with a standard deviation of 5 h 8 min, and the average duration of Work activity is 3 h 4 min with a standard deviation of 2 h 29 min.

\begin{center}
\textbf{TABLE 3}
\end{center}

%
%
%
%
%
%
%
%

The generation of the daily agenda of an individual is subject to daily activity patterns, depending on which classification this individual is categorized into. As for the majority group of people whose main activities are working, we generate their daily agenda according to the research in \citep{Bhat2000,Roorda2008,Min2008}. 6 activity patterns are adopted as shown in Table \ref{activity patterns} where ``*" stands for possible activities other than Home and Work. The percentages are taken from \citep{Min2008}, based on a survey accomplished in a China city Shangyu. In simulations, a Recreation or Medical Care activity will be generated to substitute ``*".

\begin{center}
\textbf{TABLE 4}
\end{center}

%
%
%
%
%
%
%
%
%

Once an individual has finished an activity, he/she moves by either public or personal transportation from the current location to another sublocation where the next activity is going to take place. The travel routing based on either RN or PTN can be computed according to Sec. \ref{Travel Routing}, and the required time can be also estimated taking into account the travel means, the start-point and the end-point.

Furthermore, individuals' knowledge of global epidemic situation such as the alert phases issued by WHO, and their own infection status, can influence their behavior which succeedingly is reflected in the generation of daily agenda. For example, when one is aware of the severe prevalence of some epidemic or the diagnosis of own infection, he/she probably prolongs his/her stay at home, decreases work time and avoids crowded places such as recreational SLs. Therefore, the parameters for generating activities can be tuned to adapt to different situations.


\subsubsection{Infections Due to Contacts}

People encounter and have contacts with others when conducting activities or traveling, so they possibly get infected with airborne viruses when epidemics are prevalent. Let $\sigma$ be the mean number of transmission events per hour of contact between fully infectious and fully susceptible people. For events that occur randomly in time, the number of occurrences in a period of time of length $t$ obeys a Poisson probability law with parameter $\sigma t$. Thus, the probability of no occurrences in time interval $t$ is $\textrm{e}^{\sigma t}$ and the probability of at least one occurrence is $1-\textrm{e}^{-\sigma t}$. When an infectious individual $i$ and a susceptible individual $j$ stay within a distance threshold $D^*$ (tunable for epidemics) from each other in the same sublocation for a certain period of time $T_{ij}$ (recordable in simulations), infection can occur with a probability of $P_{ij}=1-\textrm{e}^{-\sigma T_{ij}}$. According to \citep{Chowell2007,Del2007}, $\sigma$ can be estimated based on knowledge of past epidemics. For simplicity, each individual is assumed to wander around inside a sublocation during the stay and his/her accurate coordinates are obtainable. Thus the distance between two that stay in the same sublocation can be measured to assess whether it is less than $D^*$.

\subsubsection{Disease Progression}

Disease progression can be simply described by stages of (1) incubation with assumed non-infectiousness, (2) symptomatic period with infectiousness and (3) recovery/death \citep{Mei2010}. For some airborne diseases such as influenza, a susceptible individual can refrain from getting infected by vaccination. Therefore, individuals can become immunized by either natural immunization (recovery from the previous infection) or vaccination. Therefore 3 parameters of $D_\textrm{incubation}$, $D_\textrm{symptomatic}$ and $D_\textrm{vaccination}$ are introduced to our model, indicating the duration of the incubation stage, the symptomatic stage and the duration that vaccination needs to stimulate immunity, respectively. For example, we can set $D_\textrm{incubation}=1\sim 2$, $D_\textrm{symptomatic}=1\sim 7$ and $D_\textrm{vaccination}=7\sim 21$ days for influenza A (H1N1) \citep{Mei2010C}.

\section{GIS-based Implementation and Visualization}

\begin{figure}[htbp]
\begin{center}
\includegraphics[width=300 bp]{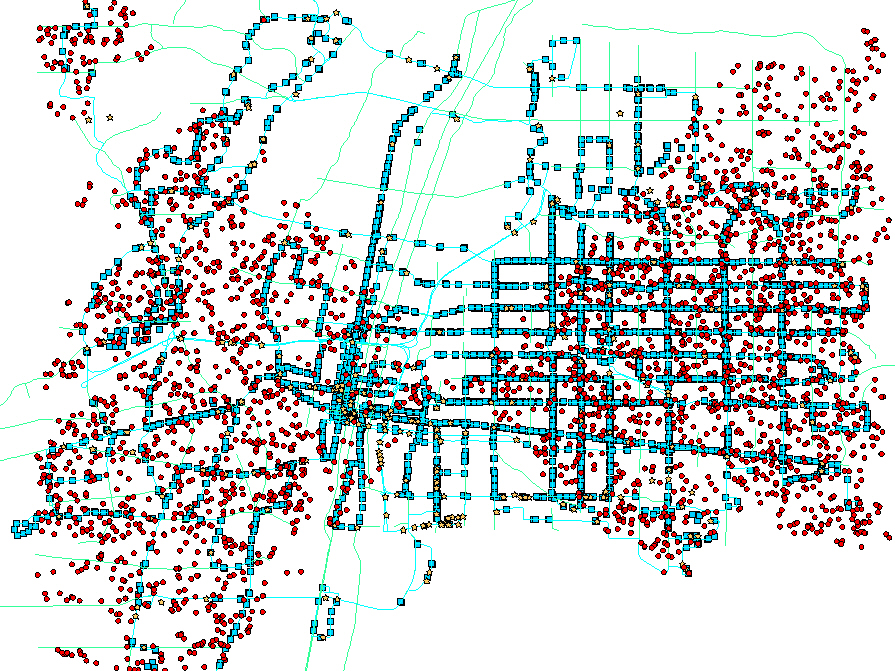}
\caption{City Visualization}\label{City Visualization}
\end{center}
\end{figure}

Based on available GIS data, we implemented the model and described into a simulation environment which can further allows for scenario testing and decision support. Fig. \ref{City Visualization} displays the visualization of a real city. Roads are shown as lines, bus/metro stops as squares, buses/metros as stars, and persons as circles. All these objects are stored, queried and manipulated efficiently based on GIS. When simulations are running, stars and circles are moving on the 2-dimensional map, representing that buses/metros operate along lines and people conduct daily activities according to their own agenda. Thus, we can record infection events and make statistical analyses of the spreading of airborne infectious diseases, e.g., locating where first infections take place, determining in which kind of sublocation (office and buses etc.) most infections occur, and illustrating the spreading situation (total patients and infection rate etc.) in each city region.

\begin{center}
\textbf{FIGURE 5}
\end{center}


%
%

\section{Conclusion}
\label{conclusion}

We have developed a novel system that integrates the most relevant geo-spatial and dynamical information required to assess the potential outbreak of airborne diseases in an urban environment. We combine GIS data with traffic and mobility patterns as well as knowledge on behavioral aspects. The information is represented in a dynamical multi-agent simulation system. The system allows for interactively exploring various alternative scenarios to support decision making and prevention, prediction and recovery of an outbreak.

In the future, our study will be focused on the parallelization and distribution of the system to support large-scale simulations. Although the system is currently being used to understand in retrospect the outbreak of influenza in a few selected densely populated small cities in China, the challenge of execution performance holds back the application of the system to simulating airborne infectious diseases in large cities with a population of, say, millions.

\section*{Competing interests}
The authors declare that they have no competing interests.

%

\section*{Acknowledgments}

This work was supported by China National Scientific Fund (No. 91024030 \& No. 91024015), ViroLab \citep{Sloot2009} and
the European DynaNets (www.dynanets.org) grant (EU Grant Agreement Number 233847). The Research was also partly sponsored by a grant from the `Leading Scientist Program' of the Government of the Russian Federation, under contract 11.G34.31.0019. In particular, we thank Dr. Viktor M\"uller and Qing Xu for their helpful suggestions.



\newcommand{\PBS}[1]{\let\temp=\\#1\let\\=\temp}
\newcolumntype{R}[1]{>{\PBS\raggedright\hspace{0pt}}m{#1}}
\newcolumntype{L}[1]{>{\PBS\raggedleft\hspace{0pt}}m{#1}}
\renewcommand{\tabularxcolumn}[1]{>{\PBS\raggedleft\hspace{0pt}}m{#1}}

\begin{tiny}
\begin{table}[htbp]
\centering \caption{Regions and Sublocations} \label{regions and sublocations}
\newlength{\LL}
\settowidth{\LL}{Scale}
\begin{tabular}{R{3cm}R{10.5cm}}\hline

\textbf{Region Type} & \textbf{Secondary Sublocation (SL) Classes} \\\hline

{Housing Region} & {\textbf{Housing SLs}: houses, apartments...; \textbf{Office SLs}: community service offices...; \textbf{Recreational SLs}: shops, community gardens...; \textbf{Classroom SLs}: children daycare...} \\\hline

{Office Region} & {\textbf{Office SLs}: offices, factory workshops...; \textbf{Recreational SLs}: cafeteria, sport places, restaurants...}
\\\hline

{School Region (for elementary, middle and high schools)} & {\textbf{Housing SLs}: faculty members' households...; \textbf{Office SLs}: teachers' offices...; \textbf{Classroom SLs}: classrooms, labs and library space...; \textbf{Recreational SLs}: cafeteria, shops, refectories...}
\\\hline

{University Region} & {\textbf{Housing SLs}: student dormitories, faculty members' households...; \textbf{Office SLs}: teachers' offices...; \textbf{Classroom SLs}: classrooms, labs and library space...; \textbf{Recreational SLs}: cafeteria, clubs, shops, refectories...} \\\hline

{Medical Region} & {\textbf{Office SLs}: doctors' offices...; \textbf{Patient Room SLs}: medical wards...; \textbf{Recreational SLs}: refectories...} \\\hline

{Recreational Region} & {\textbf{Recreational SLs}: all kinds of retail, service, meal and shop SLs} \\\hline

\end{tabular}
\end{table}
\end{tiny}

\begin{tiny}
\begin{table}[htbp]
\centering \caption{Age structure for people classification} \label{people classification}
\settowidth{\LL}{Scale}
\begin{tabular}{R{3cm}R{10cm}}\hline

\textbf{Classification} & \textbf{Description} \\\hline

{Children under 3 years} & {For children under 3 years, it is assumed that they do not have independent activities and always stay inside households.} \\\hline

{Children between 3
and 18 years} & {Their activity patterns are
assumed to be simple: go to daycare or school at school hours and stay inside households at all other times.} \\\hline

{Adults between 18
and 60 years except college students} & {They go to work at working places during day time and stay inside households during night. They visit recreational places from time to time.} \\\hline

{College students between 18
and 25} & {They go to colleagues or universities during day time and stay inside dormitories during night. They visit recreational places very often.} \\\hline

{Adults over 60 years} & {They stay around households during day time and stay inside households during night. They visit recreational places less often.} \\\hline

\end{tabular}
\end{table}
\end{tiny}

\begin{tiny}
\begin{table}[htbp]
\centering \caption{Activity types} \label{activity types}
\settowidth{\LL}{Scale}
\begin{tabular}{R{3.5cm}R{6.5cm}R{3cm}}\hline

\textbf{Activity Types} & \textbf{Places} & \textbf{Involved People Classes} \\\hline

{Work (W)} & {Office SLs, Classroom SLs (for students), Recreational SLs (for salesmen and waiters etc.), Patient Room SLs (for doctors and nurses)} & {All but children under 3 years and adults over 60 years} \\\hline

{Home (H)} & {Housing SLs}  & {All 5 classes} \\\hline

{Medical Care (M)} & {Patient Room SLs (for patients)} & {All 5 classes} \\\hline

{Recreation (R)} & {Recreational SLs (for consumers)} & {The latter 3 classes}  \\\hline

\end{tabular}
\end{table}
\end{tiny}

\begin{tiny}
\begin{table}[htbp]
\centering \caption{Activity Patterns} \label{activity patterns}
\settowidth{\LL}{Scale}
\begin{tabular}{R{4cm}R{3.5cm}}\hline

\textbf{Activity Patterns} & \textbf{Percentages(\%)} \\\hline

{HWH} & {53.4} \\

{HWH*H} &  {10.3} \\

{HW*WH} & {2.7} \\

{HWHWH} & {27.1}  \\

{HWHWH*H} & {6.5}  \\\hline

\end{tabular}
\end{table}
\end{tiny}

\newpage


\end{document}